# LINEAR POTENTIALS, AIRY WAVE PACKETS AND AIRY TRANSFORM

G. DATTOLI

ENEA - Gruppo Fisica Teorica e Matematica Applicata
Centro Ricerche Frascati, Roma

K. ZHUKOVSKY

ENEA Guest
PERMANENT ADDRESS: Faculty of Physics, Moscow State University,
Leninskie Gory, Moscow 119899 (Russia)

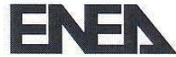

# LINEAR POTENTIALS, AIRY WAVE PACKETS AND AIRY TRANSFORM

G. DATTOLI

ENEA - Gruppo Fisica Teorica e Matematica Applicata
Centro Ricerche Frascati, Roma

K. ZHUKOVSKY

ENEA Guest
PERMANENT ADDRESS: Faculty of Physics, Moscow State University,
Leninskie Gory, Moscow 119899 (Russia)

# LINEAR POTENTIALS, AIRY WAVE PACKETS AND AIRY TRANSFORM

G. DATTOLI, K. ZHUKOVSKY

*Abstract*
*The solution of the Schrödinger equation with a linear potential is considered. We use algebraic methods to obtain the explicit form of the solution for the explicitly time dependent Hamiltonian and discuss the general conditions which allow us to get solutions in terms of the Airy functions, yielding non spreading wave packets. We analyze the relevant physical meaning of these solutions and give examples of their applications. We discuss the analogy between Airy and Gauss-Weierstrass transform.*

**Riassunto**
Nel presente lavoro si discute la soluzione dipendente dal tempo dell'equazione di Schrodinger con un potenziale lineare. Il metodo qui proposto viene utilizzato per risolvere equazioni con Hamiltoniane dipendenti esplicitamente dal tempo. Si analizzano inoltre le condizioni generali che permettono di scrivere la soluzione in termini di funzioni di Airy. Si discute il significato fisico delle soluzioni ottenute ed in particolare il ruolo dell'effetto di non diffusione del pacchetto d'onda durante l'evoluzione temporale. si mette infine in evidenza l'importanza della trasformata di Airy in tale contesto.

**Parole chiave:** equazione di Schrödinger, Fokker-Plank, trasformazione di Airy, Gauss, operatori di evoluzione, diffusione e traslazione.

# INDEX

# LINEAR POTENTIALS, AIRY WAVE PACKETS AND AIRY TRANSFORM

## 1. INTRODUCTION

The stationary solution of the Schrödinger equation with a linear potential can be expressed in terms of the Airy functions [1] The time dependent problem can be studied with the help of methods, involving various analytical means, which can be all framed within a common algebraic treatment [2]. The problem of the free propagation of an Airy packet was discussed long ago in ref. [3], where it was shown how the packet may evolve in space and time without distortion and with constant acceleration. It was also demonstrated [3] that the same effect occurred when the packet evolution was ruled by a linear (possibly time dependent) potential. Recently the problem of the Airy Packet evolution has been considered in [4]. It was shown that the most general potential, preserving the non spreading properties of the packet evolution, was a linear potential.

In the present paper we reconsider the problems, addressed in [2] – [4] with the help of the algebraic tools. We will show that, apart from a phase term, the method we use allows the factorization of the evolution operator for a linear potential in the product of two terms, representing respectively the diffusion and the translation in space. To make the analysis more general we begin with the case of a generalized form of the heat equation, including a linear term.

The problem of finding the solution of diffusive equations with a linear term playing the role of a potential is more interesting than it might be thought and it has still some aspects which have not been thoroughly explored yet. In this paper we will point out the existence of previously unnoticed features, like the link with the Airy transform and with generalized forms of polynomials of Hermite type, which make the problem worth to be reconsidered.

## 2. SOLUTION OF FOKKER-PLANK AND SCHRÖDINGER EQUATION BY THE EXPONENTIAL OPERATOR METHOD

The techniques we will adopted in the paper, will be illustrated by deriving the solution of the following generalization of the heat equation

$$\partial_t F(x,t) = \partial_x^2 F(x,t) + \beta\, x\, F(x,t) \tag{1}$$

with the initial condition:

$$F(x,0) = f(x). \tag{2}$$

Equation (1) is an evolution type equation, consisting of a diffusive part and of an additional linear contribution in the spatial coordinate. The relevant solution can be written in terms of the evolution operator $\hat{U}$ as follows:

$$F(x,t) = \hat{U} f(x), \tag{3}$$

where

$$\begin{aligned}&\hat{U} = e^{\hat{A}+\hat{B}} = e^{t\hat{H}}, \\ &\hat{A} = t\,\partial_x^2,\; \hat{B} = \beta\, t\, x,\; \hat{H} = \partial_x^2 + \beta x.\end{aligned} \tag{4}$$

The evolution operator $\hat{U}$ in (3) contains the sum of two non commuting quantities and it can be written as the ordered product of two exponential operators. Indeed, by noting that the commutation bracket between $\hat{A}$ and $\hat{B}$ satisfies the following property

$$\left[\hat{A},\hat{B}\right] = m\,\hat{A}^{\frac{1}{2}} \quad m = 2\beta\, t^{\frac{3}{2}}, \tag{5}$$

we can apply the following disentanglement identity [5]:

$$e^{\hat{A}+\hat{B}} = e^{\frac{m^2}{12}-\frac{m}{2}\hat{A}^{\frac{1}{2}}+\hat{A}}e^{\hat{B}}. \tag{6}$$

Moreover, we will make use of the following chain rule

$$e^{p\,\partial_x^2}e^{qx}g(x) = e^{pq^2}e^{qx}e^{2p\,\partial_x}e^{p\,\partial_x^2}g(x), \tag{7}$$

where $p$ and $q$ – constant parameters, and further, of its consequence [1)]

$$e^{p\,\partial_x^2}e^{qx}\cdot 1 = e^{pq^2}e^{qx}\cdot 1. \tag{8}$$

With the help of the previous identities we obtain the following form for the evolution operator, associated with the eq. (1):

$$\hat{U} = e^{\frac{1}{3}\beta^2 t^3 - \beta t^2 \partial_x + t\partial_x^2}e^{\beta t x} = e^{\Phi(x,t;\beta)}\hat{S}\hat{D}, \tag{9}$$

where

$$\hat{S} = e^{\beta t^2 \partial_x}, \tag{10}$$

$$\hat{D} = e^{t\partial_x^2}, \tag{11}$$

$$\Phi(x,t;\beta) = \frac{1}{3}\beta^2 t^3 + \beta t x. \tag{12}$$

The action of the evolution operator $\hat{U}$ on the initial condition function (2) yields the following solution of our problem:

---

[1)] The symbol $1$ in equation (8) means that the operator $e^{p\,\partial_x^2}e^{qx}$ acts on a constant, and it should be understood as $e^{p\,\partial_x^2}e^{qx}1 = e^{pq^2}e^{qx}e^{p\,\partial_x^2}1$ which reduces to the r.h.s. of (8) since $e^{p\,\partial_x^2}1 = 1$ $e^{p\,\partial_x^2}1 = 1$.

$$F(x,t) = e^{\Phi(x,t;\beta)} \hat{S} \hat{D} f(x). \tag{13}$$

We can conclude from (13) that the problem (1) with the initial condition (2) can be solved by the consequent application of the commuting $\hat{S}$ and $\hat{D}$ operators to (2), apart from the factor $e^{\Phi(x,t;\beta)}$. The explicit form of the solution (13) can now be obtained by recalling that $\hat{S}$ is in fact the translation operator:

$$e^{a\partial_x} g(x) = g(x+a) \tag{14}$$

and that the action of the diffusion operator $\hat{D}$ on the function $g(x)$ yields the solution of the ordinary heat equation, also recognized as a Gauss-Weierstrass transform (GW-T) [6]:

$$e^{b\partial_x^2} g(x) = \frac{1}{2\sqrt{\pi b}} \int_{-\infty}^{\infty} e^{-\frac{(x-\xi)^2}{4b}} g(\xi) d\xi. \tag{15}$$

Accordingly, for $a = \beta t^2$, $b=t$ we write:

$$\hat{D}f(x) \equiv e^{t\partial_x^2} f(x) \equiv f(x,t) \tag{16}$$

and

$$\hat{S}f(x,t) = f\left(x + \beta t^2, t\right). \tag{17}$$

Thus, the equation (1) with the initial condition (2) has the following explicit solution:

$$F(x,t) = e^{\Phi(x,t;\beta)} \frac{1}{2\sqrt{\pi t}} \int_{-\infty}^{\infty} d\xi \, e^{-\frac{\left(x+\beta t^2 - \xi\right)^2}{4t}} f(\xi), \tag{18}$$

provided that the integral converges.

We have found that without any assumption on the nature of the function $f(x)$ (apart the convergence of the integral), the solution of the eq. (1) can be written as follows:

$$F(x,t) = e^{\Phi(x,t;\beta)} f(x + \beta t^2, t). \tag{19}$$

The properties of this type of solutions will be further commented in the following sections. Here we just note that $f(x,t)$ is the solution of the heat equation, representing a diffusive phenomenon and, therefore, it represents a naturally spreading process.

The effect, produced by the translation operator $\hat{S}$ and the diffusion operator $\hat{D}$ is well illustrated with the example of $f(x) = e^{-x^2}$, in which the solution (18) reduces to the following:

$$F(x,t)\big|_{f(x)=\exp(-x^2)} = \frac{e^{\Phi(x,t)}}{\sqrt{1+4t}} e^{-\frac{(x+\beta t^2)^2}{1+4t}}, \quad t \geq 0. \tag{20}$$

This last result can be viewed as a generalization of the so called Gleisher rule [7], yielding the solution of the ordinary heat equation ($\beta = 0$), when the initial function is a Gaussian.

The same operational technique as employed for the solution of (1) can be exploited to solve the Schrödinger equation:

$$i\hbar \partial_t \Psi(x,t) = -\frac{\hbar^2}{2m} \partial_x^2 \Psi(x,t) + F\, x\, \Psi(x,t),$$
$$\Psi(x,0) = f(x), \tag{21}$$

where $F$ – a constant with the dimension of a force. Rescaling the variables in (21) allows us writing it in the form, similar to (1):

$$i\, \partial_\tau \Psi(x,\tau) = -\partial_x^2 \Psi(x,\tau) + b\, x\, \Psi(x,\tau), \tag{22}$$

where[2]

$$\tau = \frac{\hbar t}{2m}, \quad b = \frac{2Fm}{\hbar^2}. \tag{23}$$

Its solution can be obtained in the form of the exponential operator, analogous to that, (3) was written:

$$\Psi(x,\tau) = \hat{\overline{U}} f(x), \tag{24}$$

where

$$\hat{\overline{U}} = \exp\left(i\tau \hat{\overline{H}}\right) \quad \hat{\overline{H}} = \hat{H}\Big|_{\beta=-b} = \partial_x^2 - bx. \tag{25}$$

Following the way we solved the equation (1) and on account of the substitution $t \to i\tau$, $\beta \to -b$, we obtain

$$\hat{\overline{S}} = e^{b\tau^2 \partial_x}, \quad \hat{\overline{S}} f(x,\tau) = f\left(x + b\tau^2, \tau\right) \tag{26}$$

and

$$\hat{\overline{D}} = e^{i\tau \partial_x^2} \quad \hat{\overline{D}} = e^{i\tau \partial_x^2}, \quad \hat{\overline{D}} f(x) \equiv e^{i\tau \partial_x^2} f(x) \equiv f(x,i\tau). \tag{27}$$

Thus, the solution of our equation appears in the form of the sequence of the operators $\hat{\overline{S}}, \hat{\overline{D}}$, acting on the initial condition function:

$$\Psi(x,t) = e^{-i\Phi(x,\tau;b)} \hat{\overline{S}} \hat{\overline{D}} f(x), \tag{28}$$

where $\Phi$ is defined by (12). The integral form of the solution then writes as follows[3]:

---

[2] Note that this choice of variables implies that $\tau$ has the dimensions of a squared length, while $b$ - of an inverse cube length in such a way that $b\tau^2$ has the dimensions of a length.

$$\Psi(x,t) = e^{-i\Phi(x,\tau;b)} \frac{1}{2\sqrt{i\pi\tau}} \int_{-\infty}^{\infty} d\xi\, e^{-\frac{(x+b\tau^2-\xi)^2}{4i\tau}} f(\xi). \qquad (29)$$

Again, as well as in (19), without any assumption on the nature of the initial condition function $f(x)$ of the Schrödinger equation, its solution

$$\Psi(x,\tau) = e^{-i\Phi(x,\tau;b)} f(x+b\tau^2, i\tau), \qquad (30)$$

where $f(x,t)$ is given by (27), is expressed in term of the function of 2 variables, obtained by the consequent application of the diffusion and translation operators (27), (26) to $f(x)$.

Thus, the global phase term apart, the result of the evolution operator (25) action on $f(x)$ is the product of the combined action of the translation operator $\hat{\hat{S}}$ and the operator $\hat{\hat{D}}$, representing the evolution operator of the free particle. The explicit form of the solution can be obtained with the help of the GW-T, which allows calculation of the action of $\hat{\hat{D}}$ on the initial probability amplitude. Further discussion on this subject will be presented in the following section.

## 3. SPREADING AND NON SPREADING SOLUTIONS – AIRY PACKETS PROPAGATION

We consider now the solution (19) of the equation (1) with the initial condition, defined by the Airy function [1]:

$$f(x) = Ai\left(\frac{x}{A}\right) = \frac{1}{\pi} \int_0^{\infty} \cos\left(\frac{1}{3}\zeta^3 + \frac{x}{A}\zeta\right) d\zeta, \qquad (31)$$

where $A$ – a constant assumed to be positive. Its role is not secondary as discussed in what follows. For further convenience we introduce the generalised Airy function of two variables

---

[3)] Equation (29) has been written in complete analogy with the case, in which the "time" variable is real. In more rigorous term we should refer to the Fresnel transform, which in paraxial optics (as well as in non relativistic quantum mechanics) accounts for the wave free propagation.

$$Ai(x,z) = \hat{D} Ai\left(\frac{x}{A}\right) = e^{z \partial_x^2} Ai\left(\frac{x}{A}\right). \tag{32}$$

The function $Ai(x,z)$ is a two variable extension of the ordinary Airy function and is, on account of its definition (32), the solution of the heat equation

$$\begin{aligned} \partial_z Ai(x,z) &= \partial_x^2 Ai(x,z), \\ Ai(x,0) &= Ai\left(\frac{x}{A}\right). \end{aligned} \tag{33}$$

If $z$ is assumed to be a parameter and not a variable, it can be easily shown that $Ai(x,z)$ satisfies the following second order differential equation:

$$A^3 \frac{d^2}{dx^2} y + 2A^2 z \frac{d}{dx} y + x\, y = 0. \tag{34}$$

Furthermore, its integral representation reads

$$Ai(x,z) = \frac{1}{\pi} \int_0^\infty \cos \frac{1}{3}\left(\zeta^3 + \frac{x}{A}\zeta\right) e^{-\frac{z}{A^2}\zeta^2} d\zeta. \tag{35}$$

According to the previous remarks, the solution of (1) with the Airy function as the initial condition can be written as follows:

$$F(x,t) = e^{\Phi(x,t)} Ai(x + \beta t^2, t). \tag{36}$$

The behaviour of the function $Ai(x,z)$ for some values of $z$ is shown in Fig.1, where it is compared with the behaviour of the function $G(-x^2,z) = e^{z \partial_x^2} e^{-x^2} = \frac{1}{\sqrt{1+4z}} e^{-\frac{x^2}{1+4z}}$.

In the first case we have the pronounced damping of the amplitude oscillations without any significant spreading, while in the second case we have both – the reduction of the amplitude and the broadening of the curve.

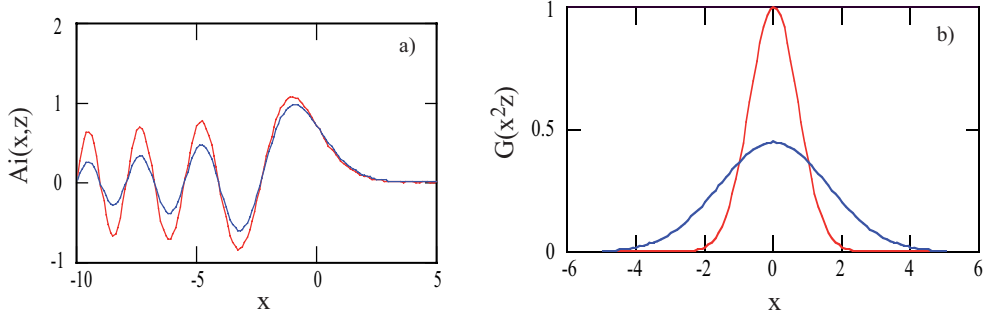

*Fig. 1 – a) The Airy function for z=0 and z=0.1; b) The Gaussian function for z=0 and z=1*

Let us now consider the quantum mechanical case of the free evolution of a particle under the action of the evolution operator (25), which for the free particle yields

$$\hat{D}\, Ai(x) = Ai(x, i\tau). \tag{37}$$

The Airy function of the complex argument $Ai(x, i\tau)$ has the following form:

$$Ai(x, i\tau) = \frac{1}{\pi}\int_0^\infty \cos\frac{1}{3}\left(\zeta^3 + \frac{x}{A}\zeta\right) e^{-i\frac{\tau}{A^2}\zeta^2} d\zeta = \frac{1}{4\pi}\int_{-\infty}^\infty \left[e^{i\varphi_+} + e^{-i\varphi_-}\right] d\zeta, \tag{38}$$

where

$$\varphi_\pm = \frac{1}{3}\zeta^3 \pm \frac{\tau}{A^2}\zeta^2 + \frac{x}{A}\zeta. \tag{39}$$

It is now important to note that $\varphi_\pm$ can be recast with regrouping the variables in the following form:

$$\varphi_\pm = \frac{1}{3}\left(\zeta \pm \frac{\tau}{A^2}\right)^3 + \left(\frac{x}{A} - \frac{\tau^2}{A^4}\right)\left(\zeta \pm \frac{\tau}{A^2}\right) \pm \left(\frac{2}{3}\frac{\tau^3}{A^6} - \frac{x\tau}{A^3}\right), \tag{40}$$

which allows for the following change of variables:

$$\zeta \to \zeta_\pm = \zeta \pm \tau \tag{41}$$

and

$$x \to \tilde{x} = \frac{x}{A} - \frac{\tau^2}{A^4}. \tag{42}$$

The rescaling of the integration variable $\zeta_\pm$ does not produce any effect since the integration runs from $-\infty$ to $+\infty$. The shift of the spatial coordinate $x$ (42) together with (41) yields the reduction of the two variable Airy function ((38) to the common Airy function (31):

$$Ai(x, i\tau) = e^{i\Theta(x,\tau)} Ai\left(\frac{A^3 x - \tau^2}{A^4}\right),$$
$$\Theta(x,\tau) = \frac{\tau}{A^6}\left(\frac{2}{3}\tau^2 - A^3 x\right). \tag{43}$$

Formula (43) confirms that free propagation does not produce any spreading as it has already been stressed in [3] and in [4].

We can also take note that the following property of the two variable Airy function

$$\hat{\tilde{D}} Ai(x,z) = Ai(x, z + i\tau) \tag{44}$$

follows from its definition itself and therefore we conclude that

$$\hat{\tilde{D}} Ai(x,z) = e^{i\Theta(x,\tau)} Ai(A^3 x - \tau^2, z). \tag{45}$$

If the Schrodinger equation under study refers to the motion of a particle, carrying the charge $q$ in the static electric field of intensity $E_0$, the force term should be specified as $F = qE_0$. With the help of the same rescaled variables (42), (41) and the assumption that the initial function is the Airy function we easily derive the solution from our previous analysis in the following form:

$$\Psi(x,\tau) = e^{-i\Phi(x,\tau;b)} Ai(x + b\tau^2, i\tau). \tag{46}$$

With the help of the formula (43) it can be reduced to common Airy functions of one variable:

$$\Psi(x,\tau) = e^{-i\Phi(x,\tau;b)} e^{i\Theta(x+b\tau^2,\tau)} Ai\left(\frac{A^3 x + (A^3 b - 1)\tau^2}{A^4}\right), \tag{47}$$

where $\Phi$ is given by eq. (12) and $\Theta$ is given by (43).

Therefore, apart from an inessential (for the purposes of the present article) global phase, we have obtained that all the dynamics of the Airy packet in the electric field is reduced to the translation of the packet, determined as follows:

$$|\Psi(x,\tau)|^2 = \left|Ai(A^3 x + (A^3 b - 1)\tau^2)\right|^2 \tag{48}$$

The role of the constant $A$ will be clarified in the concluding section.

## 4. EXPLICITLY TIME DEPENDENT POTENTIALS

Let us now consider the following more general example of the Fokker-Plank like equation:

$$\begin{aligned}
\partial_t F(x,t) &= \hat{L}(t) F(x,t), \\
\hat{L}(t) &= \alpha(t)\partial_x^2 + \beta(t) x, \\
F(x,0) &= f(x),
\end{aligned} \tag{49}$$

in which $\hat{L}(t)$ – an explicitly time dependent operator, which does not commute with itself at different times, namely

$$[\hat{L}(t), \hat{L}(t')] \neq 0. \tag{50}$$

According to (50) the equation (49) cannot be solved by simply defining the evolution operator as $\hat{U} = e^{\int_0^t \hat{L}(t')dt'}$. In addition to the ordinary disentanglement techniques, further care is necessary to include properly the time ordering contributions. To this end we should use

Feynman-Dyson time ordering methods, but the structure of the operator $\hat{L}(t)$ allows to cast the evolution operator $\hat{U}(t)$, satisfying the equation:

$$\partial_t \hat{U}(t) = \hat{L}(t)\hat{U}(t),$$
$$\hat{U}(0) = \hat{1}.$$
(51)

In the Wei Norman ordered form [5]

$$\hat{U}(t) = e^{a(t)\hat{1} + b(t)\partial_x + c(t)\partial_x^2} e^{d(t)x},$$
$$a(0) = b(0) = c(0) = d(0) = 0,$$
(52)

where the ordering functions $a,b,c,d$ are determined by the following system of differential equations [5]:

$$\dot{a} + \dot{d}\, b = 0$$
$$\dot{b} + 2\dot{d}\, c = 0$$
$$\dot{d} = \beta$$
$$\dot{c} = \alpha.$$
(53)

Its straightforward integration yields

$$c(t) = \int_0^t \alpha(t')\, dt',\quad d(t) = \int_0^t \beta(t')\, dt',$$
$$b(t) = -2\int_0^t \left[\beta(t')\int_0^{t'}\alpha(t'')\, dt''\right] dt',$$
$$a(t) = 2\int_0^t \beta(t')\left\{\int_0^{t'}\left[\beta(t'')\int_0^{t''}\alpha(t''')\, dt'''\right]dt''\right\}dt'.$$
(54)

In spite of the above written complicated form of the expression, the evolution operator can always be factorized as follows:

$$\hat{U}(t) = e^{\tilde{\Phi}(x,t)} \hat{S}(t) \hat{D}(t),$$
$$\tilde{\Phi}(x,t) = a(t) + d(t)\left[ c(t)\,d(t) + b(t) + x \right], \tag{55}$$
$$\hat{S}(t) = e^{[b(t)+2c(t)d(t)]\partial_x}, \quad \hat{D}(t) = e^{c(t)\partial_x^2}.$$

Thus we arrive to the same conclusion as before, i.e. that the solution of the Fokker-Planck equation (1) is reduced, from the operational point of view, to the combined action of the translation and the diffusion operators. Moreover, the same conclusion applies to the case of the Schrödinger equation.

## 5. DISCUSSION AND CONCLUSIONS

The various integrals, which appear in the expression for the time dependant functions (54) have definite physical meaning. Indeed, by applying the above developed method to find the solution of the Schrödinger equation with $\varepsilon = \varphi(\tau)/m$, describing the motion of an electron in a classical, time dependent electric field, we obtain the following solution[4] (apart from the phase):

$$\Psi(x,t) \propto Ai\left[ \frac{B}{\hbar^{2/3}} \left( x - \frac{B^3 t^2}{4m^2} - \int_0^t \frac{(t-\tau)}{m} \varphi(\tau) d\tau \right) \right] \tag{56}$$

where $\varphi(\tau)$ – the time dependent potential. It is the same solution as given in [2]. According to the previous equation, the time evolution of the Airy function coordinate is

$$X_c = \frac{B^3 t^2}{4m^2} + \int_0^t d\tau \frac{(t-\tau)}{m} \varphi(\tau). \tag{57}$$

---

[4] We have assumed that the initial condition for our Schrödinger equation is $\Psi(x,0) = Ai\left(\frac{Bx}{\hbar^{2/3}}\right)$, where $B$ plays the role of the constant $A$, introduced in the previous section. To clarify its meaning we consider the stationary solution of the Schrödinger equation for a charged particle in a static electric field, which can be written as $-\frac{\hbar^2}{2m}\partial_x^2 f(x) + qExf(x) = 0$. The normalization of the coordinate yields $B = \sqrt[3]{2mqE}$, so that $\frac{B}{\hbar^{2/3}}$ has the dimension of the inverse of a length. This observation clarifies the role of the constant $A$ in (31).

By taking the second derivative of $X_c$ with respect to time, we find

$$\ddot{X}_c = \frac{B^3}{2m^2} + \frac{\varphi(\tau)}{m}, \tag{58}$$

which represents the coordinate acceleration. Whereas the second term in the right hand part of (58) is clearly associated with the acceleration, induced by the electric field, the presence of the term, depending on $B$, may be surprising. As already explained in [3], its contribution is due to the fact that the Airy function is not square integrable and the centroid $\left(\langle x \rangle = \int_{-\infty}^{\infty} x f(x) dx \Big/ \int_{-\infty}^{\infty} f(x) dx \right)$ of the Airy packet cannot be defined. Thus the contribution of the term $B^3/2m^2$ in the expression for the acceleration of the coordinate does not contradict the Ehrenfest theorem.

This point may be better explained with the help of the Airy transform (Ai.-T) method [8]. The concept of the Airy transform is extremely useful, albeit not widespread known as it should be. The Ai-T of a given function is defined as

$$\tilde{Ai}(f(x)) = \Phi_\alpha(\eta) = \frac{1}{\alpha} \int_{-\infty}^{\infty} f(x) Ai\left(\frac{\eta - x}{\alpha}\right) dx. \tag{59}$$

The rules, concerning the Airy transform, follow from the operational methods we have described in the paper. We take note of the following identities:

$$\begin{aligned}
\tilde{Ai}(x f(x)) &= (\eta - \alpha^3 \partial_\eta^2) \Phi_\alpha(\eta), \\
\tilde{Ai}(\partial_x f(x)) &= \partial_\eta \Phi_\alpha(\eta),
\end{aligned} \tag{60}$$

useful to derive the Ai-T of the equation (22), which reads:

$$\begin{aligned}
i \partial_\tau \Phi_\alpha(\eta,\tau) &= -(1 + b\alpha^3) \partial_\eta^2 \Phi_\alpha(\eta,\tau) + b\eta \Phi_\alpha(\eta,\tau), \\
\tilde{Ai}(\Psi(\xi,\tau)) &= \Phi_\alpha(\eta,\tau).
\end{aligned} \tag{61}$$

The diffusive part of the above equation can be eliminated by setting $\alpha = -(1/b)^{1/3}$. This choice removes the arbitrariness of the constant $\alpha$ and allows the solution of (61) in the following simple form:

$$\Phi_{-\frac{1}{\sqrt[3]{b}}}(\eta,\tau) = e^{-i\varepsilon\tau}\Phi_{-\frac{1}{\sqrt[3]{b}}}(\eta,0). \tag{62}$$

The physical meaning of the Ai-T is perhaps all contained in the identities (60). Unlike the Fourier transform, which acts on the position operator by turning it into the derivative operator of the transformed space and the derivatives into positions, the Ai-T turns the position operator into a position operator plus a diffusive part, while the derivatives are transformed into derivatives. The momentum operator remains, therefore, the momentum operator in the transformed space, while position transforms in an unusual way and its physical meaning can be interpreted according to the authors of [2], who stated that the Airy wave packets represent a family of semi-classical orbits in phase space.

The question, regarding the meaning of the Airy transformed space, naturally rises. It depends on the choice of the transformation constant $\alpha$. The use of $\alpha = \frac{\hbar^{2/3}}{B}$ with the dimensions of a length ensures that the Airy transformed space is just the ordinary space, having $\alpha$ as characteristic length.

We do not dwell further on the physical aspects of this problem, since they have been thoroughly discussed in [2]. On the other hand we focus our attention on some mathematical consequences of the formalism, which are closely related to the properties of the Airy transform. We will just touch on these topics, which will be thoroughly investigated in a forthcoming paper.

We first note that the Ai-T preserves the structure of the Weyl algebra and we find that

$$\begin{aligned} x &\to \hat{X}_\eta = \eta - \alpha^3 \partial_\eta^2, \\ \partial_x &\to \partial_\eta, \end{aligned} \tag{63}$$

and therefore the commutation bracket, associated with derivative and position operators, remains unaffected:

$$[x, \partial_x] = [\hat{X}_\eta, \partial_\eta]. \tag{64}$$

The presence of a diffusive term in what we have interpreted as the position operator can be accounted for by means of the following operational identity [5]:

$$\hat{X}_\eta = e^{-\frac{\alpha^3}{3} \partial_\eta^3} \eta \, e^{\frac{\alpha^3}{3} \partial_\eta^3}, \tag{65}$$

which underlines the importance of higher order diffusive equations[5] for the type of problems we are dealing with.

Before proceeding further we premise the identity [1, 8]

$$e^{\frac{1}{3}u^3} = \int_{-\infty}^{\infty} e^{u\sigma} Ai(\sigma) \, d\sigma, \tag{66}$$

which will be employed in the following and which holds even if $u$ is an operator. The solution of the PDE

$$\partial_t G(x,t) = \partial_x^3 G(x,t),$$
$$G(x,0) = g(x) \tag{67}$$

is written below according to (66):

$$G(x,t) = e^{t\partial_x^3} g(x) = \int_{-\infty}^{\infty} Ai(\sigma) e^{\sqrt[3]{3t\sigma} \partial_x} g(x) \, d\sigma = \int_{-\infty}^{\infty} Ai(\sigma) g(x + \sqrt[3]{3t\sigma}) \, d\sigma \tag{68}$$

and after the variable rearrangement it reads as follows:

$$G(x,t) = \frac{1}{\sqrt[3]{3t}} \int_{-\infty}^{\infty} Ai\left(\frac{\xi - x}{\sqrt[3]{3t}}\right) g(\xi) \, d\xi. \tag{69}$$

---

[5] By higher order diffusive equations we mean the equations of the type $\partial_t S(x,t) = \partial_x^m S(x,t)$, $S(x,0) = s(x), m > 2$

In other words, we have found that the solution of the evolution problem (67) is nothing but the Ai-T of its initial condition. This result is fairly interesting for the number of reasons. For example, if the initial function is just the monomial $g(x) = x^n$, we can use the identity [9]:

$$e^{\lambda \partial_x^p} x^n = H_n^{(p)}(x, \lambda),$$

$$H_n^{(p)}(x,\lambda) = n! \sum_{r=0}^{\left[\frac{n}{p}\right]} \frac{x^{n-pr} \lambda^r}{(n-pr)! r!} \qquad (70)$$

involving the higher order Hermite polynomials $H_n^{(p)}(x,\lambda)$ [10] to conclude that the so called Airy polynomials defined by [1]:

$$ai_n(x,t) = \frac{1}{\sqrt[3]{3t}} \int_{-\infty}^{\infty} Ai\left(\frac{\xi - x}{\sqrt[3]{3t}}\right) \xi^n \, d\xi \qquad (71)$$

are nothing but the third order Hermite polynomials $H_n^{(3)}(x,t)$ and that they satisfy the recurrences [6]

$$ai_{n+1}(x,t) = (x + 3t \partial_x^2) ai_n(x,t)$$
$$\partial_x ai_n(x,t) = n \, ai_{n-1}(x,t) \qquad (72)$$

They play the role, analogous to that of the heat polynomials [11], which are associated with the ordinary heat equation and within the present context are the particular case of (70) with $p=2$. Thus, we conclude that this family of polynomials is the natural solution of the heat equation and it can be expressed by means of the following GW-T :

$$H_n^{(2)}(x,t) = \frac{1}{2\sqrt{\pi t}} \int_{-\infty}^{\infty} \exp\left[-\frac{(x-\xi)^2}{4t}\right] \xi^n d\xi. \qquad (73)$$

---

[6] These relations are the particular case of the higher order Hermite polynomials recurrences, which read

$$H_{n+1}^{(p)}(x,t) = (x + mt \partial_x^{m-1}) H_n^{(p)}(x,t)$$
$$\partial_x H_n^{(p)}(x,t) = n H_{n-1}^{(p)}(x,t)$$

Moreover, due to the properties of the GW-T we can write the following identities:

$$\begin{aligned}\tilde{H}(x\,f(x)) &= \left[\xi + 2t\,\partial_\xi\right]f(\xi),\\ \tilde{H}(\partial_x\,f(x)) &= \partial_\xi f(\xi).\end{aligned} \tag{74}$$

Thus, the analogies with the Ai-T are evident and both transforms can be framed within a more general context, having the generalized Hermite polynomials as common thread.

Concluding, we can state that the Airy polynomials represent a generalization of the ordinary Hermite family; they can also be exploited to get the associated bi-orthogonal partners to obtain the expansion of a given function in terms of Airy polynomials, as it will be shown in a forthcoming publication.